\documentclass[twocolumn,preprintnumbers,amsmath,amssymb,final]{revtex4-2}

\usepackage{graphicx}
\usepackage{dcolumn}
\usepackage{bm}

\def\BEq{\begin{equation}}
\def\EEq{\end{equation}}
\def\BEqA{\begin{eqnarray}}
\def\EEqA{\end{eqnarray}}
\def\BEn{\begin{enumerate}}
\def\EEn{\end{enumerate}}
\def\BWT{\begin{widetext}}
\def\EWT{\end{widetext}}

\def\d{\delta}

\def\s{\sigma}

\def\bra{\langle}

\def\ket{\rangle}

\begin{document}


\title{Simulating kaon mixing with Josephson phase qubits
}

\author{Andrei Galiautdinov}
\email{ag1@uga.edu}
\affiliation{
Department of Physics and Astronomy,
University of Georgia, Athens, Georgia
30602, USA
}

\date{\today}

\begin{abstract}
Superconducting circuits with Josephson junctions
distinguish themselves from other types of quantum computing architectures by
having easily controllable metastable computational states (the so-called phase qubits) with a 
very large ratio of their respective lifetimes.
In this pedagogical note I describe how we can use this remarkable property of the phase qubits to 
further test the validity of the superposition principle in the {\it macroscopic} quantum regime by
simulating the $K^0-{\bar K}^0$ mixing mechanism of particle physics.

\end{abstract}


\maketitle


{\it [NOTE: This is an old unpublished write-up, dated Jan.\ 18, 2009; submitted as is. Ideas described here motivated the analysis presented in Sec.\ V ``Tunneling Measurement'' of Ref.\ \cite{galiautdinov2012}.]}

\section{\label{sec:INTRO}Introduction}

Sometimes in physics we encounter 
systems so profoundly different that the mere thought of them having anything in common 
seems absurd.
And yet, upon closer examination, we find that under appropriate conditions both systems 
exhibit strikingly similar behaviors. The question then immediately arises: Can we use one 
of the systems to simulate the behavior of the other?

The two systems that will form the subject of this note are the neutral kaon pair 
$K^0-{\bar K}^0$ and a flux-biased Josephson junction, also known as the phase qubit.
What makes these utterly different systems so spectacularly similar is the fact that both 
have certain metastable quantum states of different energies whose lifetime ratios are of 
the same order of magnitude, $\sim 10^3$. 

Our primary goal here will be to explore the possibility of simulating the behavior of kaons 
by ``doing'' various quantum interference {\it gedanken} experiments with Josephson 
junctions. Such simulation, of course, would only be possible if the phase qubits themselves 
obeyed the basic principles of quantum mechanics, which in itself is a very nontrivial 
assumption. The reason is clear: the variables 
describing the junctions are voltages, currents, and superconducting phase differences 
that are macroscopic in nature. Will the quantization of such macroscopic quantities lead 
to the same observable effects as the usual quantization of microscopic variables
\cite{LEGGETT80}? Various experiments of the last two decades indicate that that is 
indeed the case \cite{CLARKE, PASHKIN, BERKLEY, STEFFEN}. 
Here I propose to further test the validity of quantum mechanics in the {\it macroscopic} 
regime by simulating the kaon system using such superconducting devices.

\section{The qubit Hamiltonian}

Our system, the flux biased Josephson phase qubit, consists of a superconducting loop
(typically made of Al, interrupted by a 
Josephson junction) that is driven by an external magnetic flux in such a way at to always 
keep the flowing current below its critical value, $I_0$. Under those conditions the junction 
exhibits the so-called dc-Josephson effect, in which Cooper pairs can directly tunnel through 
the insulating barrier of the junction with no need for any external voltage source, such as, 
for example, a battery \cite{FEYNMAN_LECTURES}. 

It turns out that when the system is cooled down to millikelvin temperatures, the 
superconducting phase difference $\d$ across the junction (usually viewed as a macroscopic 
{\it classical} quantity that describes how the phases of the Cooper pair wave functions on 
the two sides of the junction differ from each other) effectively decouples itself from all the 
other degrees of freedom and begins to act as a ``true'' {\it quantum} variable. Experiments 
confirm that dynamics of such macroscopic quantum variable is well represented by the Hamiltonian
 \cite{KOFMAN2006},
\BEq
\label{eq:fullH}
H = E_c p^2/\hbar^2
- E_J\cos \d +(E_J/2\beta)\left(\d - 2\pi \phi\right)^2,
\EEq
where $p \equiv (\hbar/2e)^2C\dot{\d}$ 
is the generalized momentum associated with $\d$, $C$ is the junction capacitance,
$E_c = (2e)^2/2C$ is the corresponding charging energy, 
$E_J= (\hbar/2e)I_0$ is the coupling energy (responsible for the tunneling),
$\phi= \Phi_{\rm ext}/\Phi_0$ is the external dimensionless magnetic bias flux, 
$\Phi_0= h/2e$ is the flux quantum, and
$\beta \equiv 2\pi I_0L/\Phi_0$ is the dimensionless junction inductance. 
[The reader is invited to derive this Hamiltonian from the classical equation of motion for 
the phase difference $\d$, which, in turn, can be derived 
by using the two Josephson equations, $I_J=I_0\sin\d$, 
$V = (\Phi_0/2e)\dot{\d}$, and by noticing that the total magnetic flux through the loop,
$\Phi_{\rm total} = \Phi_{\rm ext} - LI_{\rm total}$, $I_{\rm total}\equiv I_J+C\dot{V}$, 
is related to  $\d$ via $\Phi_{\rm total} = (\Phi_0/2e)(\d+2\pi n)$, $n=0,\pm 1,\pm 2, \dots$.]
When the bias has both dc and ac components,
\BEq
\phi(t) = \phi_{\rm dc} + \phi_{\rm ac} \cos(\omega_{\rm rf}t), \;
|\phi_{\rm ac}/\phi_{\rm dc}| \ll 1,
\EEq
the Hamiltonian assumes the form
\BEqA
\label{eq:Hwithdc}
H &=&
\underbrace{E_c p^2/\hbar^2 - E_J \cos \d 
+  (E_J/2\beta) \left(\d - 2\pi \phi_{\rm dc}\right)^2}_{H_{\rm dc}}
\nonumber \\
&& -(2\pi E_J\d /\beta) \phi_{\rm ac} \cos(\omega_{\rm rf}t).
\EEqA

For the purposes of quantum computing, it is convenient to describe 
junction's behavior using
the language of qubits. In that language, the relevant macroscopic quantum variables
are projected onto the so-called computational subspace spanned by the 
lowest (metastable) eigenstates 
$|0\ket$ and $|1\ket$ of $H_{\rm dc}$ that are
localized in one of the wells on the left hand side of the corrugated parabolic potential, according to
\BEq
\d \rightarrow
 -\left(\d_{11}-\d_{00}\right)\s^z/2 +  \left(\d_{11}+\d_{00}\right) I/2
 + \d_{01}\s^x.
 \EEq
 The matrix elements of the momentum are then given by
$p_{mm'}=\left(i\hbar/2E_c\right)(E_m-E_{m'})\d_{mm'}$,
which follows from the commutator $[H,\d]$ and the canonical commutation relation 
$[p, \d] = -i\hbar$.  
The resulting qubit Hamiltonian is
\BEq
\label{eq:qubitH}
H=
 -(\Delta \epsilon/2)\s^z 
 + mc^2 I
  + \hbar\Omega_x \cos(\omega_{\rm rf} t) \s^x,
\EEq
were we have introduced the qubit level splitting 
$\Delta \epsilon \equiv \epsilon_{|1\ket} - \epsilon_{|0\ket}$, the Rabi frequency 
$\Omega_x\ll \Delta \epsilon/\hbar$, and a constant term $mc^2  \ll \Delta \epsilon$, 
whose meaning will soon be clarified. 

In actual experiments with Josephson junctions the transition time 
$t_{\rm trans} \equiv 2\pi\hbar/\Delta \epsilon \sim 10^{-10}$ s is much shorter than 
the time $t_{\rm op} \sim 10^{-8}$ s to do a qubit operation. For this reason, when 
analyzing superconducting qubits, a rotating wave approximation (RWA) is typically used.

\section{Simulating $K^0-{\bar K}^0$ mixing}

\subsection{$CP$-invariant processes}

For our purposes it will be convenient to use a particular form of the RWA. We define 
it by choosing a reference bias $(\phi_{\rm dc})_{\rm ref}$ that will set
the reference splitting $\Delta\epsilon_{\rm ref} \equiv \hbar \omega_{\rm ref}$. We then 
write $\Delta\epsilon$ at arbitrary $\phi_{\rm dc}$ as
$\Delta \epsilon \equiv  \Delta \epsilon_{\rm ref} - \Delta m c^2$, $\Delta m \ll m$.
We then choose the rotating frame $e^{(i/\hbar)H_0t}(\dots)e^{-(i/\hbar)H_0t}$,
with $H_0 = -(\Delta \epsilon_{\rm ref}/2)\s^z$, and set the rf frequency to 
$\omega_{\rm rf}=\omega_{\rm ref}$. After averaging over fast oscillations we get 
the RWA Hamiltonian 
($\hbar =c = 1$)
\BEq
\label{eq:HRWA}
H_{\rm RWA} = 
\underbrace{
\begin{pmatrix}
 m_2-i/2\tau_2 & 0  \cr
0 & m_1-i/2\tau_1
 \end{pmatrix}}_{H_{{\rm st} + {\rm wk}}}
 + \underbrace{
 (\Omega_x/2) \s^x}_{H_{\rm Rabi}}.
\EEq
The term $H_{{\rm st} + {\rm wk}}$ will be used to represent a combined effect 
of the strong and weak interaction on the so-called
long- and short-lived kaon modes $|K_{2,1}\ket$ (see below), which can be 
simulated by the 0 and 1 states of the qubit,
\BEq
|K_{2}\ket \equiv |0\ket, \quad |K_{1}\ket \equiv |1\ket.
\EEq
The Rabi contribution $H_{\rm Rabi}$ will be used to model some additional processes, 
such as, for example, the influence of
the medium in which the kaons are propagating.

In the Hamiltonian above we have added the terms that take into account metastability 
of the $|0\ket$ and $|1\ket$ qubit states, and defined the new parameters $m_{2,1}\equiv m\pm \Delta m/2$ and 
$\tau_{2,1}\equiv \tau_{\rm |0\ket,|1\ket}$ corresponding to the masses and lifetimes of the $|K_{2,1}\ket$.
The mass difference $\Delta m \equiv m_2-m_1$
for actual kaons is around $3.52\times 10^{-6}$ eV, which is much smaller than their respective 
rest masses. For decays caused by the weak interaction, such as $K_{2} \rightarrow 
\pi^+ \pi^- \pi^0/\pi^0 \pi^0 \pi^0$ and $K_{1} \rightarrow  \pi^+ \pi^-/\pi^0\pi^0$,
$\tau_{2} = 0.52\times 10^{-7}$ s and $\tau_{1} = 0.89\times 10^{-10}$ s. This gives 
$\tau_{2}/\tau_{1}\sim 10^3$, which matches nicely the lifetime ratio 
$\tau_{|0\ket}/\tau_{|1\ket} $ of the corresponding computational states typically used 
in experiments with 
Josephson phase qubits.

The $|K_2\ket $ and $|K_1\ket$ states are odd and even under the combined action of 
complex conjugation and parity,
$CP|K_2\ket = -|K_2\ket$, $CP|K_1\ket = +|K_1\ket$,
which allows us to simulate the effect of $CP$ transformation (up to the imaginary $i$) 
by doing a $\pi$-rotation about the $z$-axis of the Bloch sphere,
\BEq
\label{eq:CP}
CP = -i \s^z \equiv R_z(\pi).
\EEq 
Notice that because $|K_2\ket $ and $|K_1\ket$ are the eigenstates of both $CP$ and 
$H_{{\rm st} + {\rm wk}}$,  the $CP$ transformation is a symmetry of our model Hamiltonian.

The difference in lifetimes of $|K_2\ket $ and $|K_1\ket$ leads to some remarkable 
consequences \cite{GW1}. In addition to the $CP$-states defined above, there exist 
the {\it flavor} states of definite strangeness $S$, which in our model can be swapped 
by the $CP$ transformation in Eq. (\ref{eq:CP}).
These are the famous neutral kaon modes
\BEq
\label{eq:K0}
|K^0\ket = \left( |K_1\ket + |K_2\ket \right)/\sqrt{2}, \;
|{\bar{K}}^0\ket = \left( |K_1\ket - |K_2\ket \right)/\sqrt{2}
\EEq
that are not eigenstates of $H_{{\rm st} + {\rm wk}}$ and thus have no well-defined 
masses and lifetimes. 

Imagine now that at $t=0$ the system is prepared in the pure state
$|K^0\ket$. Such preparation may be simulated by applying a strong $\pi/2$ $y$-pulse
to the ground state $|0\ket \equiv |K_2\ket$ of the qubit,
\BEq
|K^0\ket = R_y(\pi/2)|0\ket.
\EEq
Setting in Eq. (\ref{eq:HRWA}) $\Omega_x=0$ gives the following time evolution of 
the $|K^0\ket$ state under the action of
$H_{{\rm st} + {\rm wk}}$,
\BEq
\label{eq:K0t}
|K^0(t)\ket = \left(e^{-im_1t}e^{-t/2\tau_1}|K_1\ket 
+ e^{-im_2t}e^{-t/2\tau_2}|K_2\ket\right)/\sqrt{2}.
\EEq
At $t<\tau_1$, both $|K_{1,2}\ket$ modes are present in the superposition, which in 
HEP experiments is manifested by the ongoing production of pion-pion pairs due to the 
$K_{1} \rightarrow  \pi \pi$ decay. For $t\gg \tau_1$, however, after the short-lived mode 
$|K_1\ket$ had disappeared, only the slow $K_{2} \rightarrow  \pi \pi \pi$ decay is still being observed.

\begin{figure}
\includegraphics[angle=0,width=1.00\linewidth]{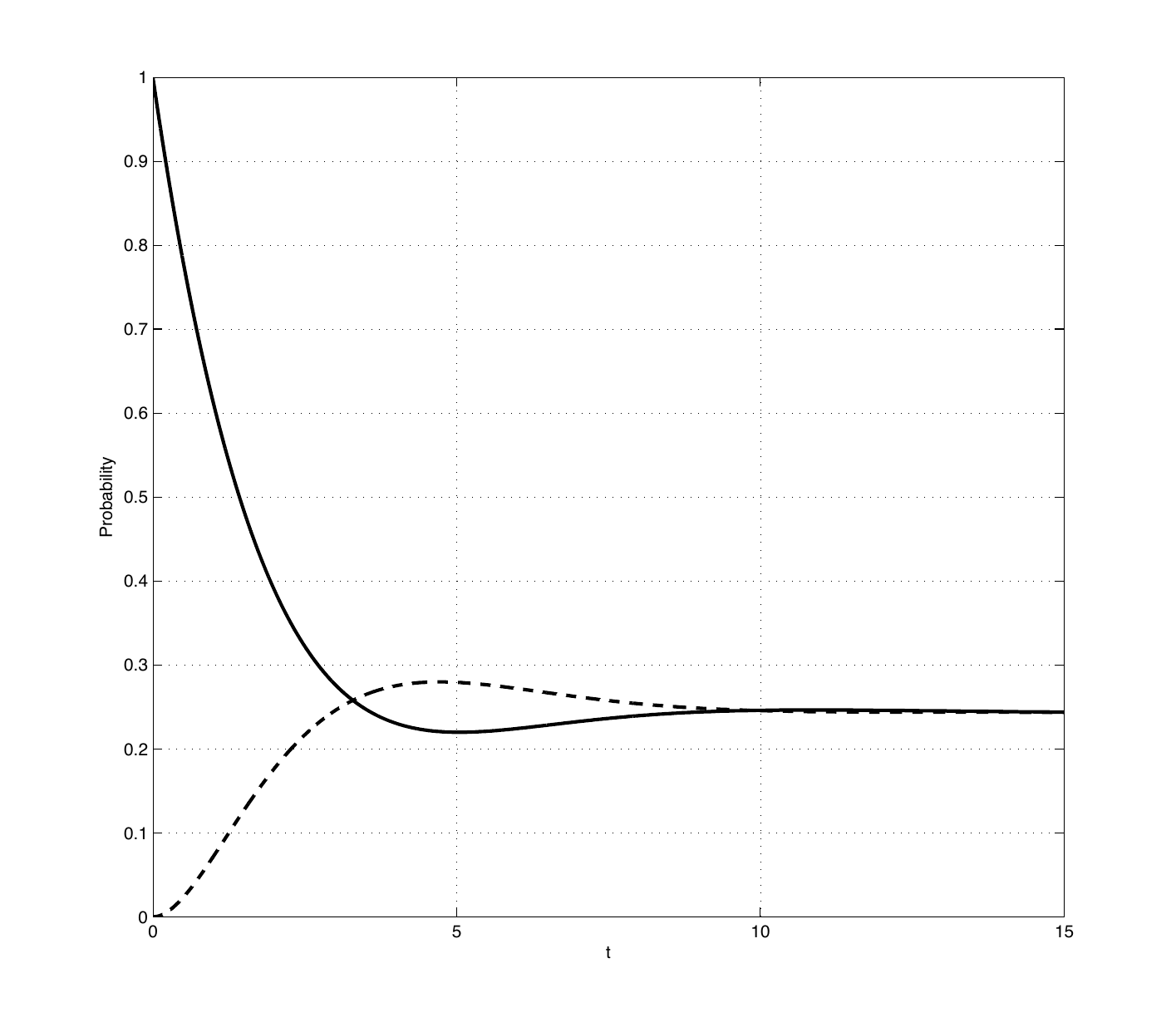}
\caption{
\label{fig:1} 
Time dependence of $P_{K^0}$ (solid curve) and $P_{{\bar K}^0}$ (dashed curve) as
given in Eq. (\ref{eq:kaonmixing}) at actually observed values $\tau_2/\tau_1 = (0.52/0.89)\times1000$
and $\tau_1\Delta m=0.477$. }
\end{figure}

One interesting phenomenon that can now be simulated with the phase qubit is the 
time-dependent mixing between the $|K^0\ket$ and $|{\bar K}^0\ket$ components
in the time-evolving state $|K^0(t)\ket$.
From Eqs. (\ref{eq:K0}) and (\ref{eq:K0t}) it follows that the probabilities of finding 
the pure states $|K^0\ket$ and $|{\bar K}^0\ket$ in the above superposition are given by
\BEq
\label{eq:kaonmixing}
P_{K^0,{\bar K}^0}(t) = \frac{e^{-t/\tau_1} + e^{-t/\tau_2}
\pm 2 e^{-(t/2\tau_1+t/2\tau_2)}\cos\left(t\Delta m \right)}{4},
\EEq
which results in a readily observable oscillation between these states of definite 
(but different) strangeness,
as depicted in Figure \ref{fig:1} (here shown at actually observed value $\tau_1 \Delta m=0.477$).

Such strangeness-changing ($\Delta S = \pm 2$) process was used 
in particle physics to experimentally determine the mass difference 
$\Delta m$ between the long- and short-lived kaon modes $K_{2}$ and $K_{1}$.
In the context of Josephson phase qubits the process can be used to test the principle of superposition 
of {\it macroscopic} quantum amplitudes in a straightforward way. 

To observe the ``kaon'' oscillations in our qubit system we have to modify
the pulse sequence to account for the fact that in the present-day experiments only 
$|0\ket$ and $|1\ket$ states can  be directly measured, not the $|1\ket \pm |0\ket$ 
states. We thus propose a new sequence that gives the same probabilities as those in
Eq. (\ref{eq:kaonmixing}),
\BWT
\BEq
\label{eq:qubitkaons}
|\psi(t)\ket := R_y(-\pi/2)U_{{\rm st} + {\rm wk}}
\underbrace{
R_y(\pi/2)|0\ket
}_{(|0\ket+|1\ket)/\sqrt{2}}
= \frac{e^{-im_1t}e^{-t/2\tau_1} + e^{-im_2t}e^{-t/2\tau_2}}{\sqrt{2}} |0\ket 
+
\frac{e^{-im_1t}e^{-t/2\tau_1} - e^{-im_2t}e^{-t/2\tau_2}}{\sqrt{2}} |1\ket.
\EEq
\EWT

\begin{figure}
\includegraphics[angle=0,width=1.00\linewidth]{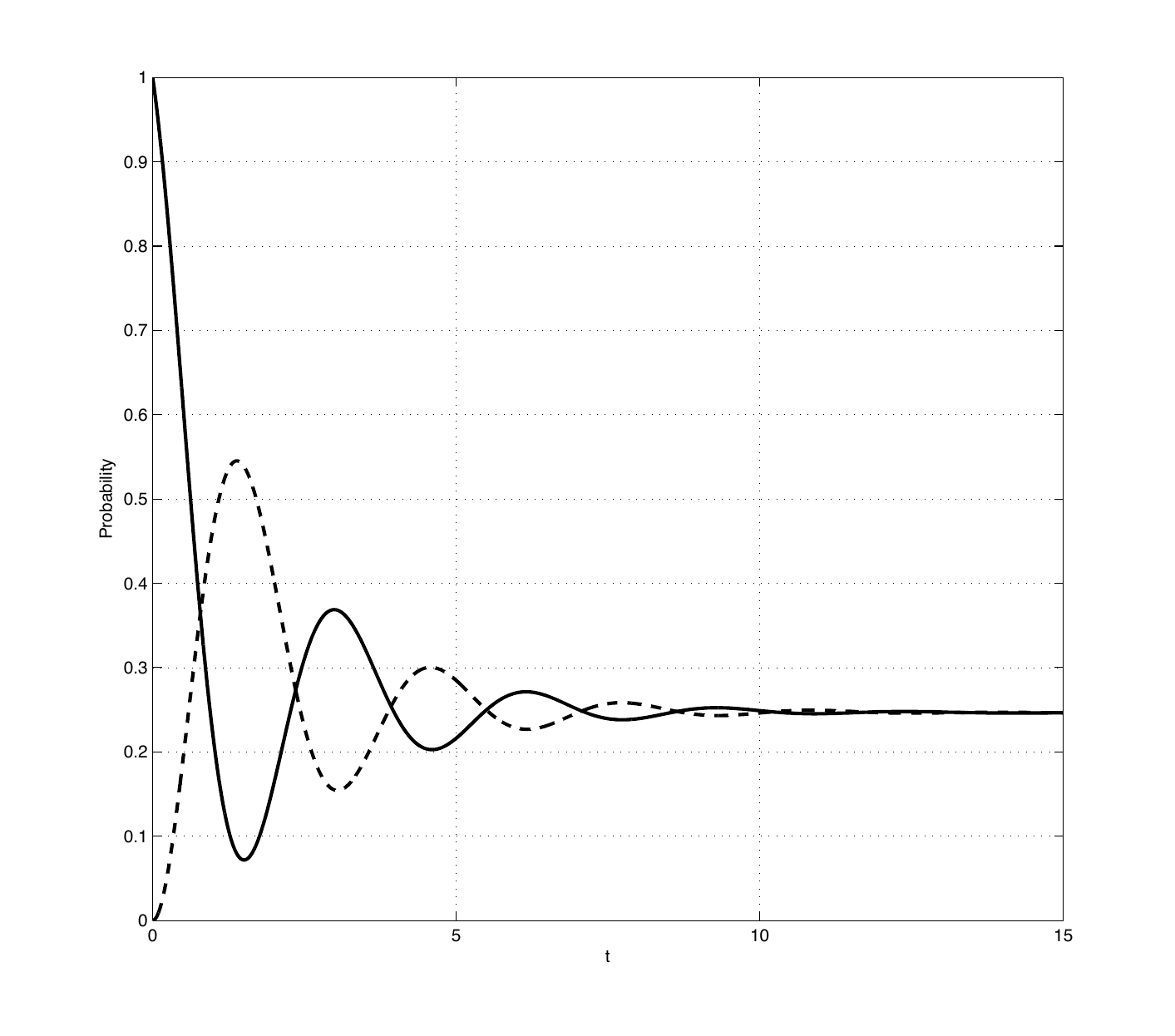}
\caption{
\label{fig:2} 
Predicted time dependence of probabilities $P_{|0\ket}$ (solid curve) and $P_{|1\ket}$ 
(dashed curve) 
for a single phase qubit control sequence given in Eq. (\ref{eq:qubitkaons}) at 
$\tau_2/\tau_1 = 1000$, $\tau_1\Delta m=2.000$. }
\end{figure}

The role of the final $R_y(-\pi/2)$ rotation is to change the $|0\ket$ and $|1\ket$ states into
the corresponding superpositions $|0\ket\mp|1\ket$ that differ from each other by a minus sign, 
the extra sign that shows up in the second term on the right hand side of Eq. (\ref{eq:qubitkaons}), 
which is responsible for the oscillating interference pattern. 
Figure \ref{fig:2} shows how the probabilities of the qubit states 
 would vary in time for $\tau_2/\tau_1=1000$, $\tau_1 \Delta m=2.000$, if the superposition 
principle were {\it macroscopically} valid.

An even more interesting version of the above experiment can be performed with two 
capacitively coupled qubits. The interference would be observed between the two excited 
states $|01\ket$ and $|10\ket$, {\it detuned} by
$\Delta m \equiv E_{|01\ket}-E_{|10\ket}> 0$ and having lifetimes $\tau_{01} < \tau_{10}$. 
The proposed sequence is
\BEq
\label{eq:2qubitkaons}
|\psi(t)\ket :=
U_{\rm int} U_{\rm det}(t)
 \underbrace{U_{\rm int}R_y^1(\pi)|00\ket
}_{ (|01\ket -i |10\ket)/\sqrt{2}},
\EEq
where $U_{\rm det}(t)$ is a free evolution with the noninteracting detuned Hamiltonian,
\BEq
H_{\rm det} = 
\begin{pmatrix}
E_{10} -i/2\tau_{10} & 0\cr 
0 & 0
\end{pmatrix}_{2} 
+ 
\begin{pmatrix}
 E_{01}-i/2\tau_{01} & 0\cr 
0 & 0
\end{pmatrix}_{1},
\EEq
 and 
$U_{\rm int}$ is the {\it resonant} evolution for
a time $t_{\rm int} \equiv \pi/4g$ 
with the coupling Hamiltonian,
\BEq
H_{\rm int}= (g/2)(\s^x_2\s^x_1+\s^y_2\s^y_1).
\EEq
If the coupling is sufficiently strong we may ignore the tunneling effects during the 
entangling operations. In that case, the probabilities of
measuring $|10\ket$ and $|01\ket$ states are again given by 
Eq. (\ref{eq:kaonmixing}) and Figure \ref{fig:2}.
Notice that the role of the last $U_{\rm int}$ operation here is similar to that of the 
final $R_y(-\pi/2)$ rotation in the single-qubit case: to ``rotate'' the $|01\ket$ and $|10\ket$
states in such a way as to produce the extra minus sign responsible for their interference.

Another process that can be simulated with a single phase qubit is the so-called 
{\it regeneration phenomenon}. Starting again at $t=0$ in the $|K^0\ket$ state, we 
first wait for a time $t_1$, such that $\tau_2 \gg t_1 \gg \tau_1$, until the short-lived 
component vanishes and
the state becomes $|K^0(t_1)\ket = e^{-im_2t_1}e^{-t_1/2\tau_2}|K_2\ket/\sqrt{2}$;
see Eq. (\ref{eq:K0t}).
We then imagine that this ``beam'' of kaons enters a thin slab of matter
with which it briefly interacts in some way for $t_2\ll t_1$, which may be modeled 
by the full Hamiltonian Eq. (\ref{eq:HRWA}) containing the Rabi term. 
The state will then evolve into a new superposition
$|K^0(t_1+t_2)\ket \simeq e^{-im_2t_1}e^{-t_1/2\tau_2}(C_1(t_2)|K_1\ket
+ C_2(t_2) |K_2\ket)$, where the short-lived mode had reappeared, which is the essence 
of the regeneration phenomenon. 

\subsection{$CP$-violating processes}

In 1964 it was discovered \cite{CG1982} that the
long-lived kaons decayed into two charged 
pions, a process forbidden by the 
$CP$ symmetry of $H_{{\rm st} + {\rm wk}}$. 
\begin{figure}
\includegraphics[angle=0,width=1.00\linewidth]{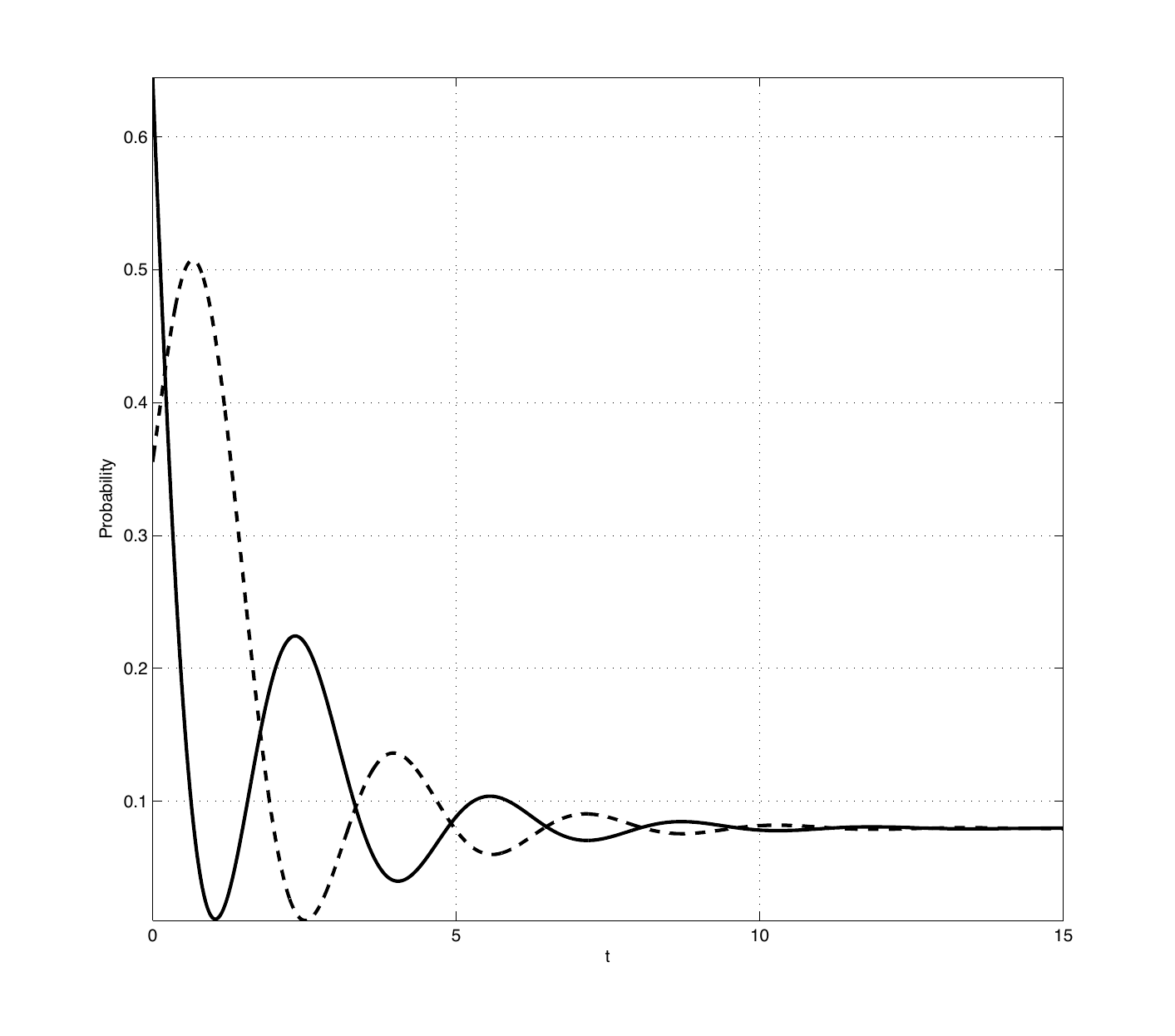}
\caption{
\label{fig:3} 
Predicted time dependence of probabilities $P_{|0\ket}$ (solid curve) and $P_{|1\ket}$ 
(dashed curve) 
for a single phase qubit control sequence given in Eq. (\ref{eq:qubitkaons}) at 
$\tau_2/\tau_1 = 1000$, $\tau_1\Delta m=2.000$,
that starts with $|K_L\ket$, $\varepsilon = 0.525(1+i)$.  }
\end{figure}
The true states describing the long- and short-lived kaon modes turned out to be 
coherent superpositions
of the $|K_2\ket$ and $|K_1\ket$ states defined above,
\BEq
|K_L\ket = \frac{|K_2\ket + \varepsilon |K_1\ket}{\sqrt{1+|\varepsilon|^2}}, \;
|K_S\ket = \frac{\varepsilon |K_2\ket + |K_1\ket}{\sqrt{1+|\varepsilon|^2}},
\EEq 
where $\varepsilon = 1.621(1+i)\times 10^{-3}$ is a small $CP$ violation parameter
\cite{GW1}.
These modes are neither the eigenstates of $H_{{\rm st} + {\rm wk}}$, nor are they
orthogonal to each other, $\bra K_S|K_L\ket \neq 0$. 
We can simulate various interference effects due to non-vanishing $\varepsilon$ by 
``creating'' such modes with simple qubit rotations by an angle 
$\theta = 2\arctan |\varepsilon|$:
\BEq
|K_L\ket = e^{i\theta(\s^x-\s^y)/2\sqrt{2}}|0\ket, \;
|K_S\ket = e^{i\theta(\s^x+\s^y)/2\sqrt{2}}|1\ket.
\EEq
 Figure \ref{fig:3} simulates the sequence in Eq. (\ref{eq:qubitkaons})
 that starts in the $CP$-violated state $|K_L\ket$, $\varepsilon = 0.525(1+i)$, instead 
of the ground state $|0\ket$. Notice how at this special value of $\varepsilon$ each of 
the two probabilities 
 vanish at first and then reappear again.

\section{Conclusion}
\label{sec:conclusion}

In summary, I have shown how metastability of the computational states in a flux biased 
Josephson phase qubit can be used to simulate the kaon mixing mechanism that would 
allow a direct verification of the superposition principle in the {\it macroscopic} quantum 
regime. I have also described single- and two-qubit control sequences that can be implemented 
using currently available superconducting quantum computing architectures.

\begin{acknowledgments}

This work was supported by IARPA 
under grant W911NF-08-1-0336 and by the NSF under grant CMS- 
0404031. 

\end{acknowledgments}

\end{document}